\documentclass[
 aps,
 pre,
 amsmath,amssymb,
 reprint,
]{revtex4-1}
\usepackage{graphicx}
\usepackage{siunitx}
\usepackage{url}
\usepackage{epstopdf}

\begin{document}

\title{Statistical analysis of phase formation in 2D colloidal systems}

\author{Hauke Carstensen$^{\ast}$}
\author{Vassilios Kapaklis}
\author{Max Wolff}
\affiliation{Department of Physics and Astronomy, Box 516, SE-751 20 Uppsala, Sweden}
\date{\today}

\begin{abstract}%
Colloidal systems offer unique opportunities for the study of phase formation and structure since their characteristic length scales are accessible to visible light.
As a model system the two dimensional assembly of colloidal magnetic and non-magnetic particles dispersed in a ferrofluid (FF) matrix is studied by transmission optical microscopy.
We present a method to statistically evaluate images with thousands of particles and map phases by extraction of local variables.
Different lattice structures and long-range connected branching chains are observed, when tuning the effective magnetic interaction and varying particle ratios.
\end{abstract}
\maketitle
\footnotetext{\textit{Department of Physics and Astronomy, Uppsala University, Box 516, SE-~75120~Uppsala, Sweden. E-mail: hauke.carstensen@physics.uu.se}}

\section{Introduction}
Phase behavior and phase transitions have been studied in colloidal systems, examples of such studies focusing on electrical~\cite{Yethiraj} or magnetic interaction~\cite{alert2014landscape,yan2012linking} have been published recently.
Of particular interest are studies on frustration~\cite{PhysRevLett.116.038303}, network formation~\cite{ranmohotti2013salt, demortiere2014self, PhysRevE.93.030601, PhysRevE.95.032606}, crystallization~\cite{pham2016crystallization, de2015entropy, castillo2015universality, morimoto2000cluster, ebert2008local} or the glass transition~\cite{PhysRevLett.116.098302}.
The beauty of microscopy in this context is that the structures are imaged in real space and can be directly visualized.
This allows to calculate directly the free energy of a system from the configuration of the colloidal particles and compare it to theoretical prediction.
However, real thermodynamic statements can only be made for very large numbers of particles approaching the thermodynamic limit.\\
One way to overcome this challenge are scattering methods directly probing the ensemble average in Fourier space.
From the transformation into real space then pair correlation functions~\cite{0305-4470-8-5-004} can be extracted and compared to theory.
However, the Fourier transform is rather abstract concept and less straight forward than optical microscopy, because the phase problem results in non-unique solutions.\\
Here, we present an alternative approach, by directly evaluating microscope images.
By collecting a large number of images and stitching them together combined with automatic particle detection we are able to access thermodynamic quantities and order parameters.\\
As an example we study the self assembly in a two dimensional system with two types of particles, magnetic and non-magnetic micro-beads.
The beads are dispersed in a ferrofluid (FF), which gives both types of particles an effective magnetic behavior~\cite{erb2009magnetic,carstensen2015phase}.
By varying the FF concentration the interaction between the particles can be tuned. For different particles interactions, ratios and densities a variety of structures can be observed~\cite{khalil2012binary}.
Statistical analysis is particularly powerful for systems with many local minima in the free energy, which can result in meta-stable phases or frustration hindering phase transitions.
From the statistical analysis the respective structures can not only be identified but their extension and number can also be linked to the energy landscape.\\
From the applied point of view, the understanding of the formation of branching chains~\cite{klokkenburg2006quantitative}, for example, is important for the understanding of magneto-rheological fluids, in which the chain formation alters the viscosity of the fluid or the assembly of lattices resulting in colloidal crystals.

\section{Experimental section}
The colloidal system, magnetic and non-magnetic micro beads dispersed in ferrofluid, is confined by two glass slides, which are separated by a 25 $\mu$m spacer sealing the samples. The microbeads, which were obtained from Microparticles GmbH, are polystyrene beads with a diameter of 10 $\mu m$, where the magnetic ones are coated with a shell of magnetic nanoparticles. The ferrofluid is a stable dispersion of magnetic iron oxide nanoparticles with a diameter of 10 nm in water, purchased from LiquidResearch.
Samples with different ferrofluid concentrations are analyzed by transmission light microscopy while an out-of-plane magnetic field with a field strength of 5~mT was applied.
For each sample a larger set of images is taken in scanning mode.
The sample stage is moved by stepper motors (Trinamic PD42-1-1141-TMCL) in a snake-like pattern.
Within a time of approximately four hours, 220~images are taken from each sample and stitched together.\\
The particle positions and the bead types are extracted by image analysis.
From this data the local particle density and composition can be associated to each bead and for each bead the coordinates and characteristic variables are saved in a list file.
In a next step this information can be used for statistical analysis and as an example the self assembled structure around each bead is analyzed, e.g. depending on the local density or composition.
Crystalline ordering can be found by detection of characteristic angles and particle distances.
For more disordered phases branching chain structures are described by the number of connected beads in one cluster and the coordination numbers of each bead.

\subsection{Stitching}
\begin{figure}[ht]
	\centering
		\includegraphics[width=0.40\textwidth]{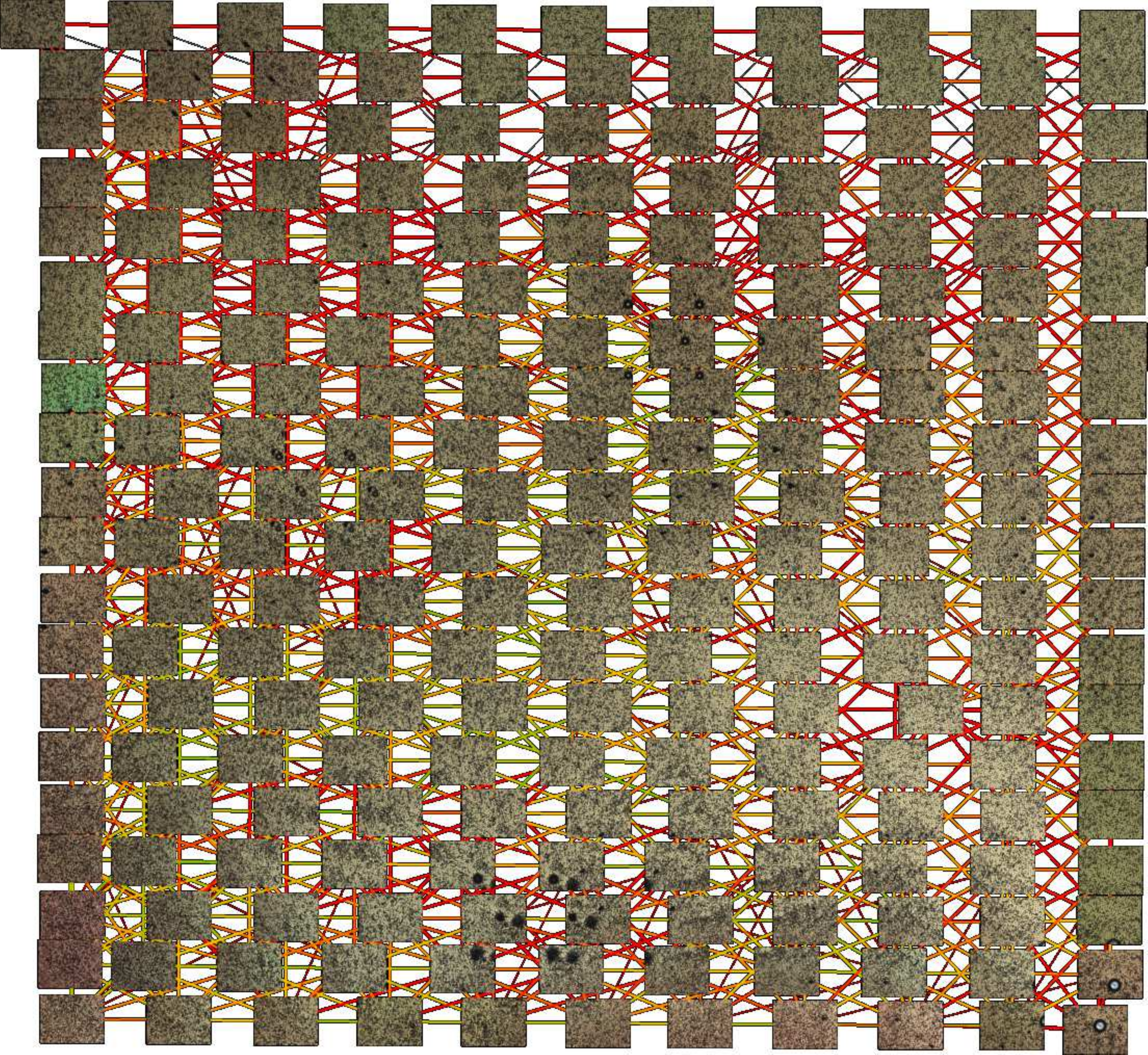}
	\caption{Hugin image stitching: The image shows the positions of the microscope pictures and the correlations between overlapping images (green is strong, yellow medium and red is weak correlation.)}
	\label{fig:layout}
\end{figure}

\begin{figure}[ht]
	\centering
		\includegraphics[width=0.50\textwidth]{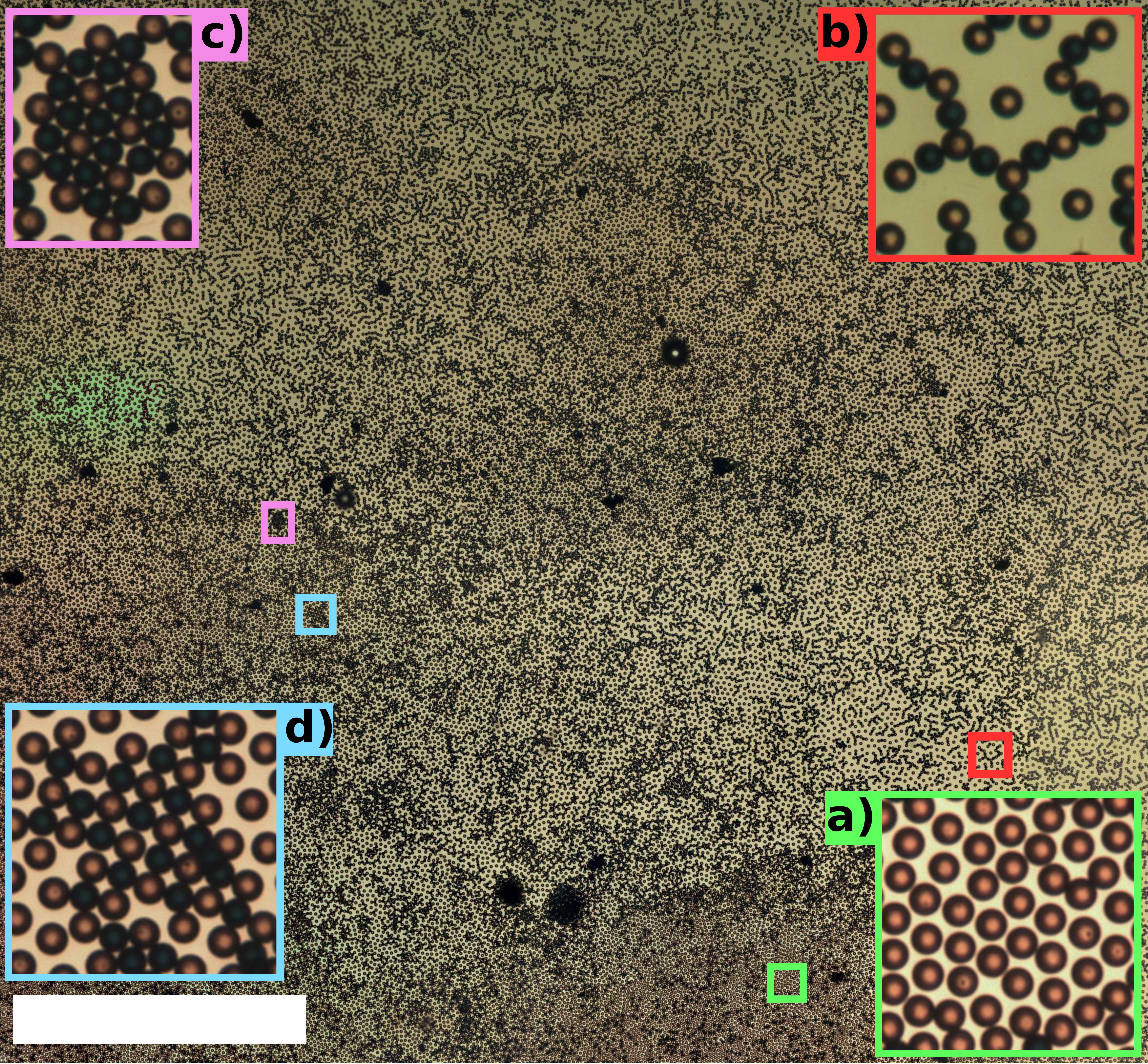}
	\caption{The stitched image shows around $3.5\times3.5$ mm$^2$ of the sample at a resolution of around 500~megapixel. Different bead arrangements can be observed, depending on density and composition: (a) Isolated beads, (b) branching chains, (c) hexagonal (honeycomb) lattice, (d) cubic lattice. The scale bar is 1 mm.}
	\label{fig:stitched}
\end{figure}

Images from each sample are stitched together to one map after they have been taken in a snake like pattern. The stitching is done with the open source software Hugin~\cite{hugin}.
Each image has a rough coordinate from where it was taken. However, due to the limited precision of the mechanical components, these coordinates are not sufficient for stitching.
The dimension of each image are $4000\times3000$ pixel and two images next to each other have an overlap of around 50 \% of the image area.
Features, high contrast points, are detected and compared for all pairs of images, where overlaps are expected based on the stepper motor coordinates.
Based on the best matching the exact coordinates are calculated.
Figure~\ref{fig:layout} displays the layout of the single images and their correlations.\\
Additionally, the exposure time and white balance is adjusted based on the overlap. A high resolution stitched map is exported.
Figure \ref{fig:stitched} shows a picture stitched from 220 single images.
\subsection{Image analysis}
The beads are identified automatically based on their circular shape by the Hough algorithm in Matlab~\cite{imfindcircles} using the fact that the gradient vectors on the circle circumference intersect at the circle center.
The non-magnetic beads are marked with a blue dye and are distinguished digitally. For each of the stitched images all bead positions are extracted. The image is processed in overlapping sections, where each section is a fraction of the large image. This has the advantage of saving memory and for each bead the section, in which it was found, is stored. The information about the sections can be used to speed up further processing of the data, because the search for neighboring beads can be limited to pairs of beads in the same or in neighboring sectors. The computation time grows linear with the number of beads N, instead of growing with N$^2$, if all possible pairs of beads are analyzed.

\section{Results}
\begin{figure*}[htb]
	\centering
		\includegraphics[width=1\textwidth]{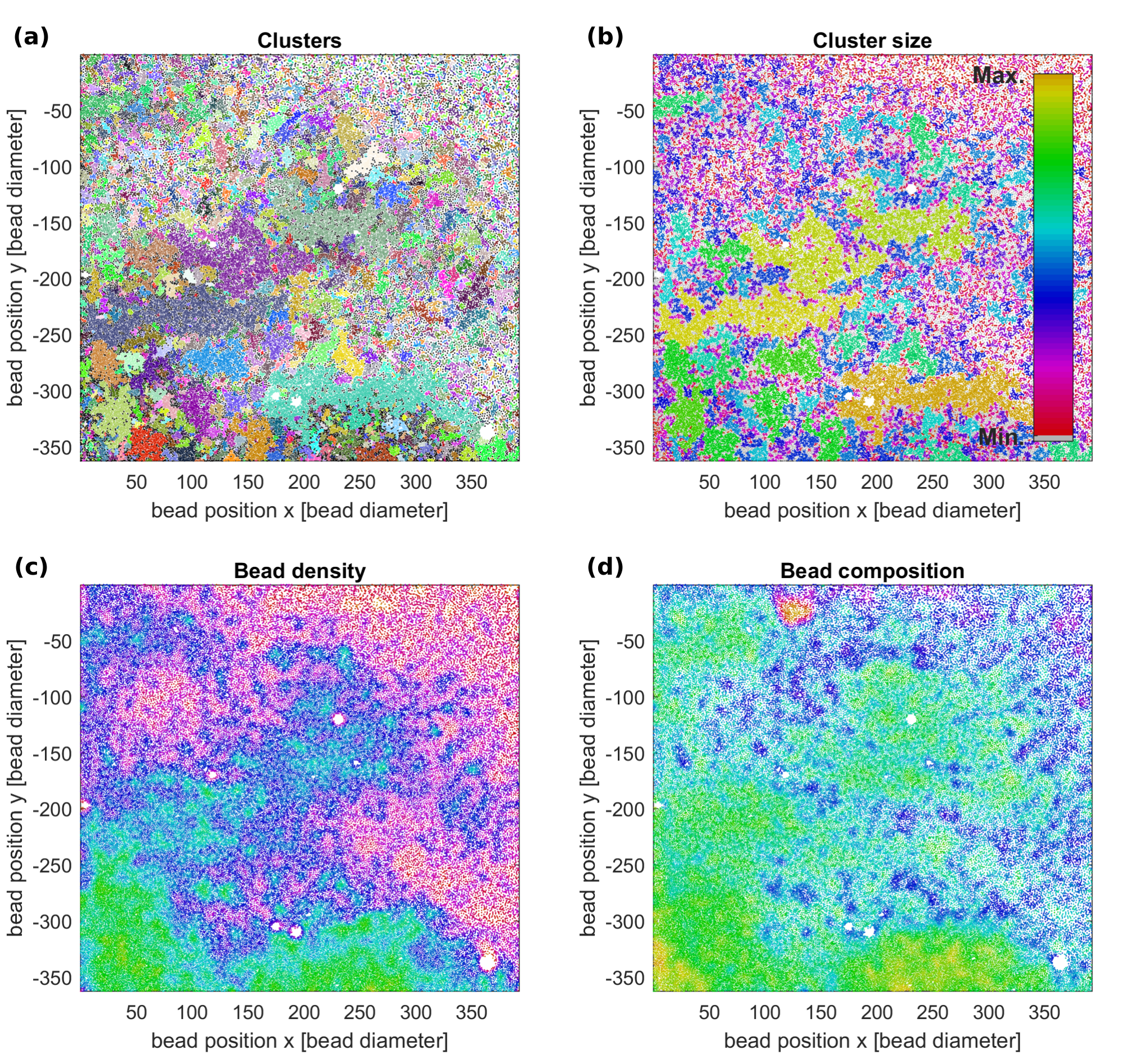}
	\caption{Local variables: 
	(a) Clusters: Every cluster of connected beads is displayed in a different (random) color. Isolated beads are drawn in black. 
	(b) Clustersize: The bead color represents the logarithmic cluster size. Here, gray color represents isolated beads.
	(c) Bead density: The color represents the density of beads. 
	(d) Bead composition: Each bead is colored depending on the fraction of magnetic beads in its vicinity.
	The label on the axis are in units of bead diameter, where 1 diameter is 10 $\mu$m. A total number of 80.000 beads is detected.}
	\label{fig:S2}
\end{figure*}

\begin{figure*}[htb]
	\centering
\includegraphics[width=1\textwidth]{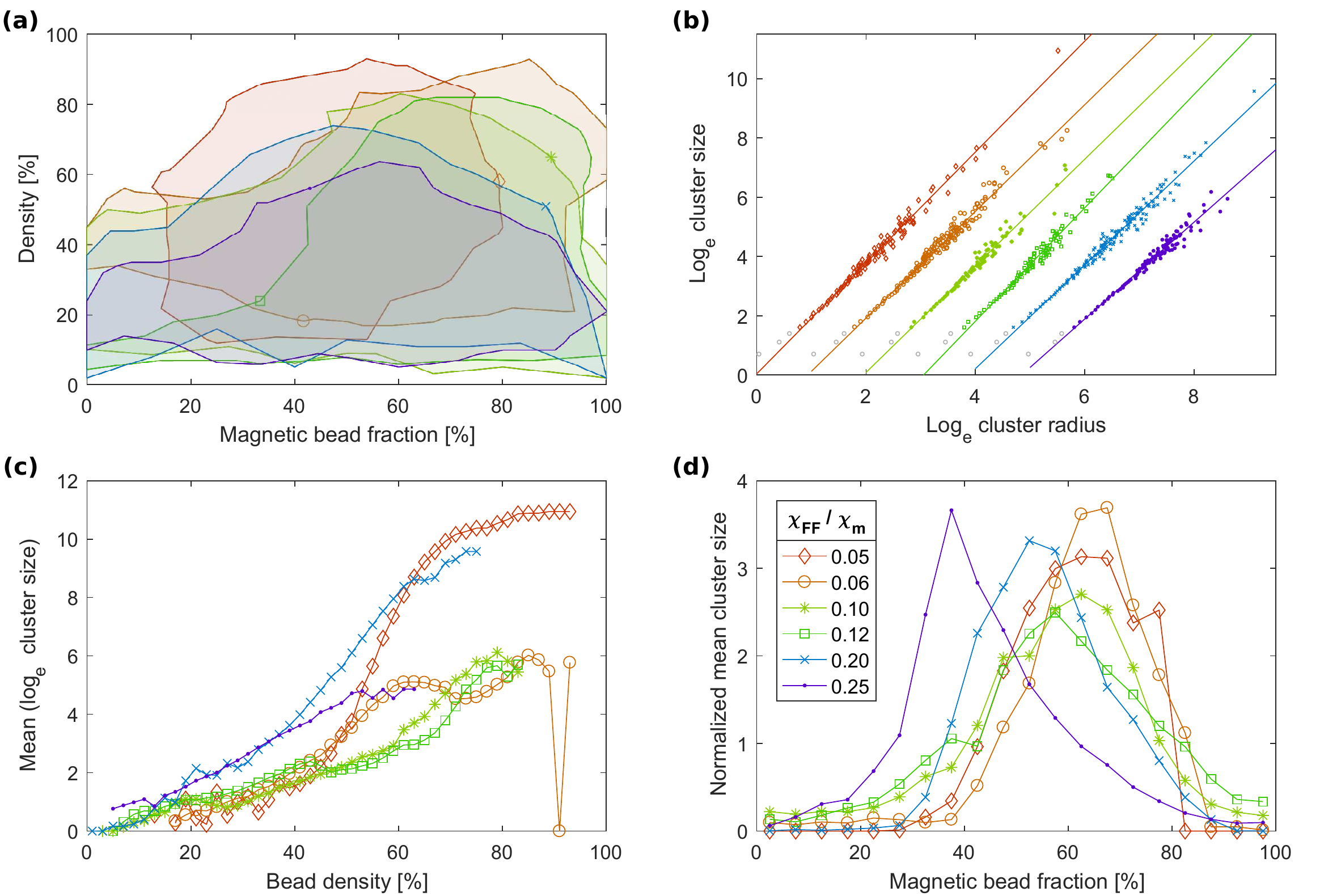}
	\caption{
	(a) Parameter spaces: For each sample the drawn area contains the point cloud of density and composition of the individual beads. 
	(b) Log-log plot of the number of beads in each cluster vs the radius of gyration of the cluster in units of bead radii: The fractal dimension is the slope of the linear regression (plotted as line) of all clusters that consists of 5 or more beads. Smaller cluster are drawn as gray circles. The data for $\chi_\text{FF} / \chi_\text{m}=$ 0.05 to 0.25 have increasing offsets on the x-axis to improve readability.
	(c) Clustersize vs bead density: The cluster size is the mean value of all beads with the same density.
	(d) Cluster size vs fraction of magnetic beads: Only beads with a density under 0.6 are evaluated. The curves are normalized by the average cluster size of each sample.}
	\label{fig:data}
\end{figure*}

Six samples were prepared with different FF susceptibilities $\chi_\text{FF} / \chi_\text{m}=$ 0.05 to 0.25, where $\chi_\text{m}$ is the susceptibility of the magnetic beads.
The dependency of the cluster size on bead density and composition, which is the fraction of magnetic beads, is analysed. Because density and composition are not homogeneous and vary within each sample, they are described as local variables and are calculated for each bead by counting neighboring beads within a threshold radius of 5 bead diameter. Direct neighbors are defined by a distance between their centers of less then 1.05 times their diameters.
All beads that are connected by a path of direct neighbors form a cluster, the number of beads in each cluster and its size are extracted.\\
For the sample with $\chi_\text{FF} / \chi_\text{m}=0.06$ the individual clusters, the cluster size, the bead density and composition are shown in figure~\ref{fig:S2}.
Noticeably, the cluster size and the bead composition correlate with the bead density. Fig. \ref{fig:data}~a shows for each sample the area that contains the density/composition point cloud. While all but one sample represent the whole range of bead composition, the bead density only reaches from a few percent to around 80\% and for densities over 60\% the bead composition is biased. The average cluster size for varying densities is shown in Fig. \ref{fig:data}~c. The cluster size depends exponentially on the density for densities under 60\%. Theoretically, it goes to infinity as the density goes to the maximum packing density $\rho_\text{max}\approx0.91$. However, as described previously this only holds in the thermodynamic limit but for a system like the present one, density fluctuations on larger length scales and the image boundaries lead to finite cluster sizes.\\
The cluster size depends also on the bead ratio (magnetic to non-magnetic) as shown in fig.~\ref{fig:data}~d. Only beads with densities under 60 \% are taken into account. For higher values the bead composition is strongly biased (as shown in \ref{fig:data}~a), especially because the cluster size depends critically on the density. Each point in the graph is the mean value of all data points with the respective composition. The largest cluster size is found for samples with a mixture of magnetic and non-magnetic beads and the composition depends on the ferrofluid susceptibility. 

The fractal dimension can be calculated as $d=\log(N)/log(R_\text{gyr})$, where $N$ is the number of beads in one cluster and $R_\text{gyr}$ is the radius of gyration of the cluster. Figure~\ref{fig:data}~b shows the number of beads vs the diameter for each cluster in a log-log plot.
The slope of a linear fit through all clusters (above a threshold size of 5 beads) is the fractal dimension, the results are varying in the range $d=$ 1.63 to 1.91 between the samples.

\section{Discussion}
The self-assembly of the beads is the result of the magnetic dipole interactions between the beads, which depend on their magnetic moments. The moments of the beads are out-of-plane. For two beads of the same type, the moments are parallel to each other resulting in a repulsive force between the beads. For two beads of unlike types the moments are anti-parallel to each other and the dipole-dipole force is attractive. The competition between attractive and repulsive forces determines which structures can be assembled.
The magnitude of the dipole-dipole force between two beads with moments $m_1$ and $m_2$ is 
\begin{align}
F\propto m_1 m_2
\end{align}
The behavior of the microbeads in the ferrofluid can be described by effective magnetic moments $m_i$, which depend on the susceptibility of the ferrofluid~\cite{erb2009magnetic}\\
\begin{equation}
m_i=3\frac{\chi_i-\chi_\text{FF}}{\chi_i+2\chi_\text{FF}+3}VH,
\end{equation}
where $\chi_i$ is the susceptibility of the beads ($\chi_i\approx 0$ for the non-magnetic type), $\chi_\text{FF}$ is the susceptibility of the ferrofluid, V the volume of the beads and H is the applied magnetic field.\\
With increasing ferrofluid susceptibility, the magnetic moments of the magnetic beads $m_\text{m}$ decrease. The effective moments of the non-magnetic beads $m_\text{n}$ are negative, which means they are anti-parallel to the external field, and the amplitude of moments is increasing with the ferrofluid susceptibility.

With increasing FF susceptibility, the repulsive forces between like beads decrease for the magnetic beads and increase for the non-magnetic ones. The attractive force between two unlike beads, $F\propto m_\text{m} m_\text{n}$, increases with the FF susceptibility (the maximum is at $\chi_\text{FF} / \chi_\text{m}\approx 0.5$).\\
A simple case of an assembled structure is a chain, which is one dimensional and built from alternating magnetic and non-magnetic beads, because consecutive like bead would repulse each other. Therefore, 50~\% of the beads in a chains are magnetic ones.
A variety of two dimensional lattice structures has been reported with increasing fraction of magnetic beads as the ferrofluid susceptibility was increased~\cite{khalil2012binary}.\\
In the colloid presented here, the structures that are assembled are patchy clusters and branching chains. With a fractal dimension of around 1.7 the structures are closer to 2D clusters. Structures assembled in in-plane magnetic fields were reported of having lower fractal dimensions~\cite{byrom2013magnetic}. The magnetic bead fraction that leads to the largest cluster growth has a peak that shifts with increasing FF susceptibility, as shown in fig.~\ref{fig:data}~d. From the minimization of magnetostatic energy (as in the 2D case~\cite{khalil2012binary}) it would be expected that for increasing FF susceptibility the peak shifts to higher fractions of magnetic beads. However, the opposite is observed and the peak shifts to lower fractions of magnetic beads. A possible explanation is that the zeta-potential of the non-magnetic beads is much lower than that of the magnetic beads (14~mV compared to 48~mV), which makes them more sticky. Therefore, the non-magnetic beads need to have higher effective magnetic moments until the magnetic interaction is strong enough to move them.

\section{Conclusion}
We present a method to automatically process large microscope images. This approach allows to limit finite size artifacts and allows the statistical analysis of long range particle correlation. As an example we study magnetic and non-magnetic particles solved in a FF of different concentrations and, therefore, different particle interactions. Within each sample the particle density and composition are fluctuating and the parameters have to be described as local variables rather than by the average over the whole sample. This fluctuations result in the assembly of different structures within one sample, of which only a small fraction appears in each single microscope image.\\
Because the structures are only weakly ordered, statistical methods are the best choice to extract quantitative information. 
The statistical approach allows to describe and quantify ordering and structure in partially ordered systems. Colloids with micro-beads typically form such partially ordered states, because the thermal energy is low, compared to the particle interactions. One example are magneto-rheological fluids, in which the formation of chains increases the viscosity, when a magnetic field is applied.

\section{Acknowledgements}
We thank Niklas Johansson and Anders Olsson, for constructing parts of the experimental setup.
We acknowledge financial support from STINT (contract number: IG2011-2067), Swedish research council (contract number: A0505501), the Carl Tryggers Stiftelse (Contract number: CT 13:513) as well as ABB group.
  
%\bibliography{paper} %your .bib file
%\bibliographystyle{unsrtnat} %the RSC's .bst file

\end{document}